# Establishing Nationwide Power System Vulnerability Index across US Counties Using Interpretable Machine Learning

Junwei Ma[1*], Bo Li[1], Olufemi A. Omitaomu[2], Ali Mostafavi[3]

[1] Ph.D. Student. Urban Resilience.AI Lab, Zachry Department of Civil and Environmental Engineering, Texas A&M University, College Station, Texas, United States.

[2] Distinguished R&D Staff. Oak Ridge National Laboratory, Oak Ridge, Tennessee, United States.

[3] Professor. Urban Resilience.AI Lab, Zachry Department of Civil and Environmental Engineering, Texas A&M University, College Station, Texas, United States.

[*] Corresponding author: Junwei Ma, E-mail: jwma@tamu.edu.

## Abstract

Power outages have become increasingly frequent, intense, and prolonged in the US due to climate change, aging electrical grids, and rising energy demand. However, largely due to the absence of granular spatiotemporal outage data, we lack data-driven evidence and analytics-based metrics to quantify power system vulnerability. This limitation has hindered the ability to effectively evaluate and address vulnerability to power outages in US communities. Here, we collected ~179 million power outage records at 15-minute intervals across 3022 US contiguous counties (96.15% of the area) from 2014 to 2023. We developed a power system vulnerability assessment framework based on three dimensions (intensity, frequency, and duration) and applied interpretable machine learning models (XGBoost and SHAP) to compute Power System Vulnerability Index (PSVI) at the county level. Our analysis reveals a consistent increase in power system vulnerability over the past decade. We identified 318 counties across 45 states as hotspots for high power system vulnerability, particularly in the West Coast (California and Washington), the East Coast (Florida and the Northeast area), the Great Lakes megalopolis (Chicago-Detroit metropolitan areas), and the Gulf of Mexico (Texas). Heterogeneity analysis indicates that urban counties, counties with interconnected grids, and states with high solar generation exhibit significantly higher vulnerability. Our results highlight the significance of the proposed PSVI for evaluating the vulnerability of communities to power outages. The findings underscore the widespread and pervasive impact of power outages across the country and offer crucial insights to support infrastructure operators, policymakers, and emergency managers in formulating policies and programs aimed at enhancing the resilience of the US power infrastructure.

# Introduction

Electric power systems serve as critical lifelines that underpin modern societies and enable the functioning of nearly every aspect of contemporary existence [1]. However, with the increasing global climate change, various extreme natural disasters, such as hurricanes and heat waves, are threatening the resilience of power systems [2-4]. In the US, between 2018 and 2020, Hurricane Florence, Michael, Laura, Sally, and Delta collectively caused severe outages that affected 0.6 to 4.3 million customers at their peak [1]. Similarly, 2021 Winter Storm Uri caused widespread power outages that impacted 25 states and over 150 million Americans [5]. In addition to natural disasters, power outages also result from various incidents, such as electrical component failures, supply shortages, physical attacks, vandalism, cyberattacks, and wildlife interference [6]. The increasingly frequent, intense, and prolonged power outages are disrupting transportation, communications, water supply, and healthcare systems, thus seriously undermining public well-being [7-9]. As a result, effectively evaluating and addressing vulnerability of power systems in communities has become an urgent priority.

Power system vulnerability refers to the susceptibility of a power system to potential harm, affecting the extent to which community members and their assets are exposed to power outages. Recent studies have increasingly emphasized the significance of assessing power system vulnerability to mitigate social suffering and formulate policies for promoting power infrastructure resilience, particularly by examining the extent of power outages. For example, Flores, McBrien et al. (2023) analyzed the social vulnerability during the 2021 Winter Storm Uri by examining power outage distribution, duration, and sociodemographic disparities related to these outages [5]. Feng, Ouyang et al. (2022) explored the compound risk of tropical cyclone- and heatwave-induced power outages in Harris County, Texas to examine how the risk evolves with changing climate and propose strategies to enhance the resilience of coastal power systems [3]. Sayarshad and Ghorbanloo (2023) evaluated the resilience of power line outages caused by wildfires in Sonoma County, US, aiming to help utilities design more resilient power lines in wildfire-prone areas [10]. However, these case-based analyses focus on the short-term impacts of isolated extreme weather-induced events on power systems. With the rising frequency and intensity of such disruptions caused by global climate change [11], it has become crucial to assess long-term and large-scale patterns and consequences. In addition to extreme weather events, various daily power outages stem from electrical component failures, supply shortages, physical attacks, vandalism, cyberattacks, and wildlife interference [6]. These frequent but localized outages also significantly affect human life but are often overlooked in research. Moreover, prior studies limited the geographic scope to specific regions in the US. For example, Dugan, Byles et al. (2023) developed an index to quantify social vulnerability to prolonged power outages using census tracts in Colorado as a case study [12]; Ganz, Duan et al. (2023) analyzed power outage data from eight major Atlantic hurricanes between 2017 and 2020 to assess the impact of hurricanes on nine southeastern US states [13]. Flores, Northrop et al. (2024) collected outage data from non-New York City urban and rural areas to evaluate the lagged effect of severe weather on power outages [14]. Given the economic and social disparities across US communities, a nationwide assessment of power outages is essential to investigate the heterogeneity of power system vulnerability across various geospatial contexts (e.g., urban vs. rural, power system operators, and regions with varying energy structures).

A data-driven characterization of power system vulnerability hinges on examining historical power outage patterns at scale. However, a major obstacle in conducting such a comprehensive analysis has been the lack of publicly available power outage data at a large scale with proper spatiotemporal resolution. Hanna and Marqusee (2022) highlight that the absence of extensive datasets impedes the examination of complex interactions between long-duration outages and system vulnerability [15]. In an era of increasingly severe outages, despite the lack of such outage data, there is an urgent need for a large-scale, high-resolution, and nationwide assessment of power system vulnerability to inform mitigation plans and policies.

Another gap associated with data availability and granularity issues is the lack of reliable and generalizable metrics for assessing power system vulnerability. Previous studies have proposed various metrics to measure outage characteristics. For example, some researchers quantified outage extent by considering the period during which customers without power exceed a specific threshold [2,14]. Flores, McBrien et al. (2023) introduced the concept of "power-out person-time", an interaction between outage duration and the number of affected customers [5]. Some studies rely on metrics from the electricity engineering field to measure outage impacts, such as SAIDI (System Average Interruption Duration Index) and SAIFI (System Average Interruption Frequency Index) [16-18]. However, the extent of power outages cannot be solely captured by a single metric. The integration of a combined set of metrics and their interaction are essential to properly quantify and evaluate the extent of vulnerability in the power systems of a region. The characterization and quantification of vulnerability has been done in the context of different socio-technical systems (such as the CDC/ATSDR social vulnerability index [19] and socio-economic-infrastructure vulnerability index [20]) and has shown to be very effective in informing plans and policies. Yet, such data-driven integrated index is direly missing for power system vulnerability.

Recognizing these gaps, this study aims to construct a comprehensive system of metrics based on granular historical power outage data and use it in creating a machine learning-based index that captures the full extent of vulnerability to power outages. To achieve this, we retrieved ~179 million power outage records at 15-minute intervals from the Environment for Analysis of Geo-Located Energy Information (EAGLE-I[TM]) platform operated by Oak Ridge National Laboratory (ORNL) [21]. This large-scale and high-resolution power outage data covers 3022 counties (96.15% of the US continental areas) between November 2014 and December 2023. By applying a 0.1% threshold to screen out the non-valid outage records, we identified a total of 3,022,915 power outage events. Drawing inspiration from environmental hazards exposure models [22-24], we developed a systematic framework to assess power system vulnerability based on three dimensions: frequency, duration, and intensity. For each dimension, multiple features were created to enable spatial and temporal analysis of outage trends across the US (Fig 1). Using these features, we trained and validated interpretable machine learning models, specifically XGBoost combined with SHAP, to examine the non-linear relationship among the features in distinguishing counties in terms of their power system vulnerability, and accordingly to reveal the relative importance of the features. Subsequently, a PSVI system with power system vulnerability values, scores, and ratings was computed through the multiplication of the features by their corresponding weights (Fig 6). Based on the computed PSVI for each county and across the decade, we further characterized the following spatiotemporal patterns: (1) spatial hotspots of power system vulnerability; (2) temporal trends and areas with growing extent of power system vulnerability; (3) variations across urban vs. rural areas and the effect of form and structure characteristics on the extent of power system vulnerability; (4) variations of vulnerability across regional transmission organizations; and (5) association between renewable energy sources and power system vulnerability. The results depict an alarming picture of how ubiquitous and widespread power outages have been across the US counties and offer valuable support of data, metrics, and methodology for infrastructure operators, policymakers, and community leaders to guide the development of policies and programs aimed at strengthening the resilience of the US power infrastructure.

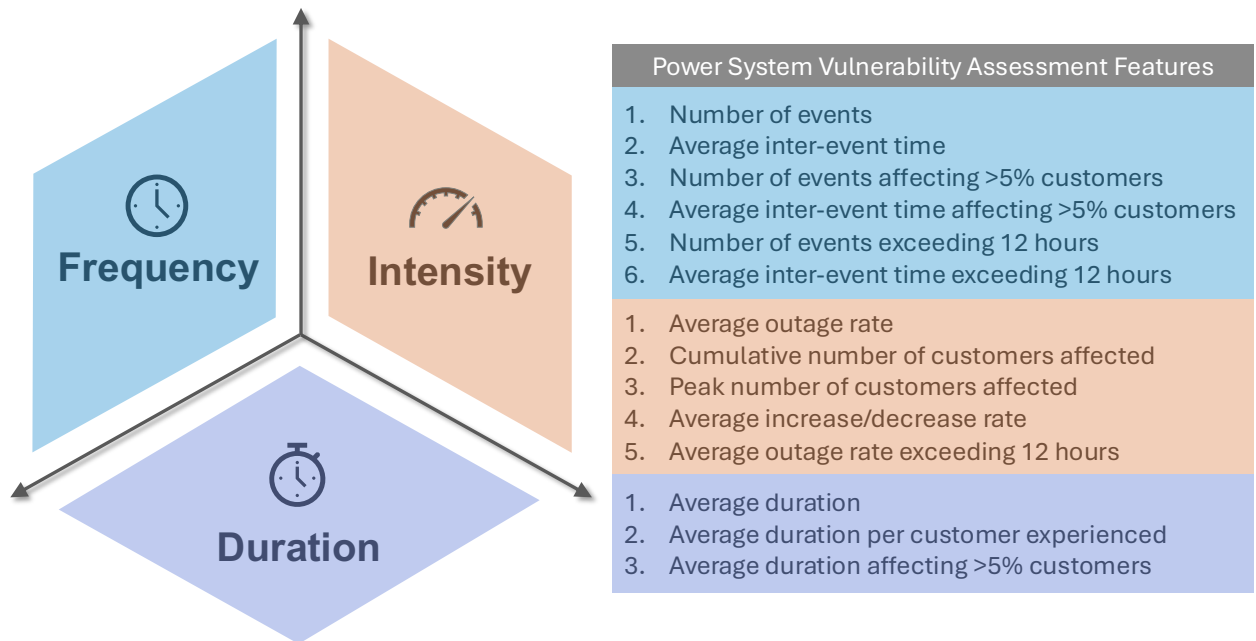

**Fig 1. Conceptual framework for assessing power system vulnerability.** We proposed a power system vulnerability assessment framework that captures three dimensions: frequency (6 features), intensity (5 features), and duration (3 features). See Methods for further detail.

# Results

**Power system vulnerability modeling by XGBoost and SHAP**

From November 2014 to December 2023, we collected a total of 179,053,397 power outage records at 15-minute intervals across 3022 US counties. During this period, the accumulative user outage time reached 7.86 billion user-hours, affecting approximately 31.47 billion customers (Supplementary Fig 1-3). These figures underscore the widespread disruption to customer service and the vulnerability of US power infrastructure. Notably, the coastal areas and Great Lakes megalopolis experienced more severe outages. Such significant geographic variation in power outage exposure highlights the dire need for integrated nationwide metrics and quantitative examinations of spatial disparities for power system vulnerability.

To systematically capture the characteristics of power outage extent, we developed a power system vulnerability assessment framework. This framework involves 14 key features across three dimensions: frequency, intensity, and duration (Fig 1). During the study period, US counties experienced an average of 1002.3 power outage events, with an average outage rate of 1.5% to total customers. The average annual outage duration was 7.3 days, meaning that each county experienced outages for approximately 2.0% of the year. The average interval between power outage events is 7.1 days, indicating that the counties experienced a power outage event approximately every week. Over the past decade, power outage events cumulatively affected 540,915 customers in each county on average. The average peak ratio of affected customers to total customers was 52.8%, and the monthly fluctuation in the number of power outage events reached 501.5%. These features highlight the widespread and substantial power outages in the US, which could cause a significant impact on daily life and economic activities. Definitions and further statistics of the features are described in the Methods section, while the spatial distribution maps are shown in Supplementary Fig 4.

To create an integrated metric, we utilized the interpretable XGBoost-SHAP model to identify the key factors influencing power system vulnerability and to calculate the PSVI. XGBoost is a powerful decision tree-based algorithm that improves model performance by iteratively generating new trees from the initial weak learners [25]. In this study, we framed the problem as a binary supervised classification task, where the 14 features served as explanatory variables, and the National Risk Index (NRI) obtained from FEMA [26] served as the dependent variable (see Methods for further detail). The NRI provides a comprehensive assessment of overall risk for US counties by integrating multiple hazards and vulnerability factors into a singular metric. By training the XGBoost model to predict the NRI based on our power system vulnerability features, we effectively gauged the extent to which power system vulnerability features explain variations in the county-level risk. The XGBoost model exhibited strong out-of-sample performance, with the F1 score of 0.7937, accuracy of 0.7835, precision of 0.8051, recall of 0.7826, and AUC-ROC of 0.8955 (Fig 2f). We also evaluated eight additional machine learning models, and the XGBoost outperformed all of them (Supplementary Table 3).

To obtain the relative importance of each feature, we used SHapley Additive exPlanations (SHAP). SHAP is an interpretable machine learning model that quantifies the contribution of each feature to individual predictions [27]. A SHAP value greater than 0 indicates a positive contribution of a feature to the mean prediction of the dependent variable and vice versa (see Methods for further detail). Fig 2g illustrates the relationship and relative importance of the 14 features driving power system vulnerability. Notably, three inter-event time features contribute negatively to power system vulnerability: average inter-event time (-0.048), average inter-event time affecting >5% customers (-0.021), and average inter-event time exceeding 12 hours (-0.001), suggesting that longer intervals between power outage events reduce both their frequency and overall vulnerability. The other features show positive contributions, with the cumulative number of customers affected having the highest importance (+0.516), followed by the average duration per customer experienced (+0.289) and average outage rate (+0.062). In addition, we grouped the 14 features into the three dimensions and calculated their cumulative importance (Fig 2g, pie chart). The intensity dimension contributed the most (52.7%), followed by duration (30.1%) and frequency (17.2%). The intensity-related features, such as average outage rate, cumulative number of customers affected, peak number of customers affected, average increase/decrease rate, and average outage rate exceeding 12 hours, played a most significant role in determining power system vulnerability. The frequency dimension ranked lowest, largely due to the inclusion of the three negatively contributing features. Overall, these 14 features, distributed across the three dimensions, provide a comprehensive characteristics of power outage events and form the basis of our PSVI.

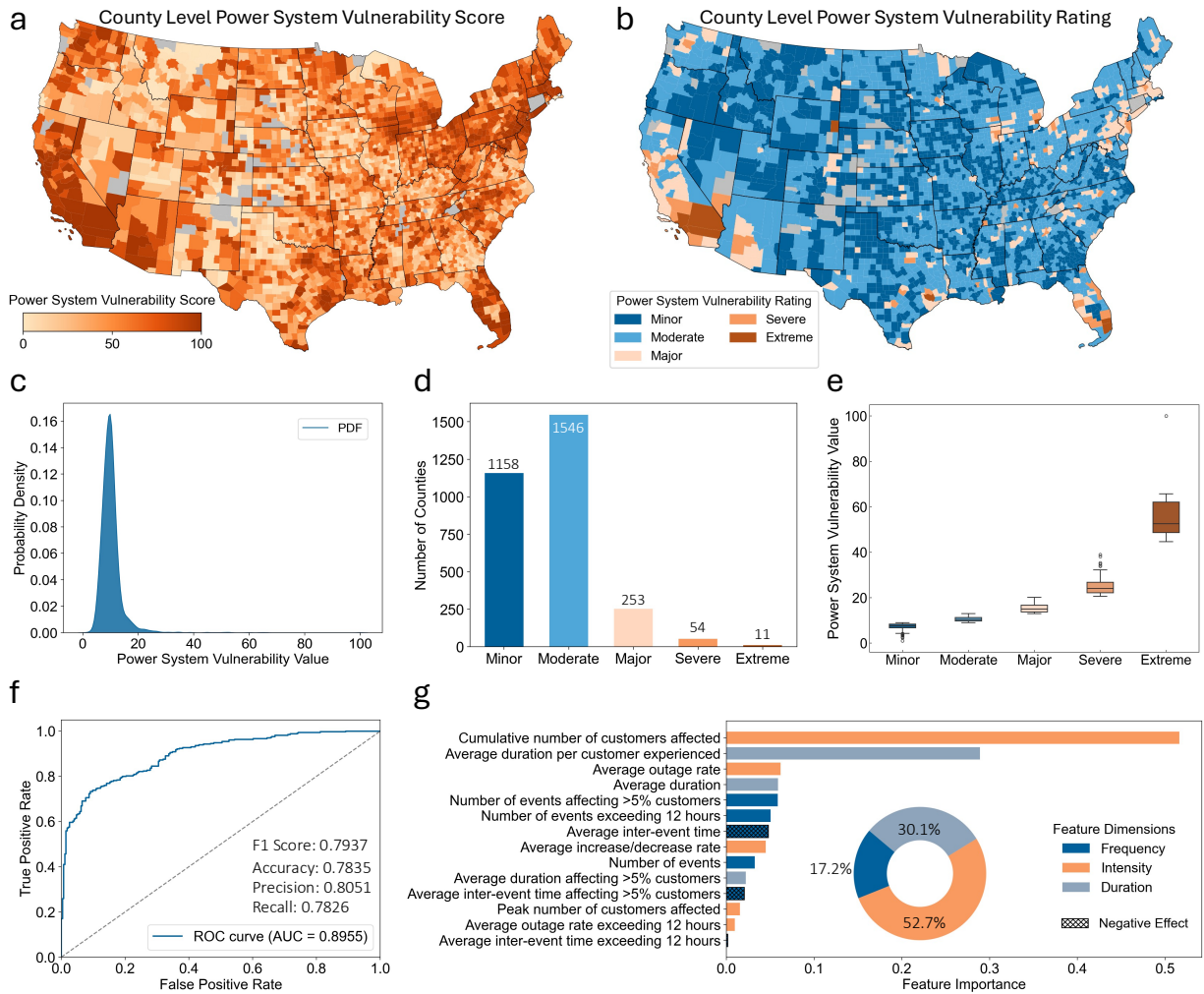

**Fig 2. Descriptive statistics of the decadal power system vulnerability index. a.** Spatial distribution of the power system vulnerability scores at the county level; **b.** Spatial distribution of the power system vulnerability ratings at the county level; **c.** Probability density curve of the power system vulnerability values; **d.** Bar plot of the number of counties across the five categories of the power system vulnerability ratings; **e.** Boxplot of the five categories of the power system vulnerability ratings. The Kruskal-Wallis H test confirmed the significant differences among the categories ($p<0.001$); **f.** AUC-ROC curve and performance indicators of the XGBoost model. **g.** SHAP importance distribution for the 14 features and 3 dimensions. Maps cover 3022 counties, with gray areas representing counties without data.

**Decadal distribution of power system vulnerability**

Using the feature importance results derived from the XGBoost and SHAP models, we developed the PSVI which comprises three components: value, score, and rating. The value represents the actual power system vulnerability. The score ranks each county on a scale from 0 to 100 based on its percentile relative to all other counties. The rating is a qualitative classification, using K-means clustering algorithm to categorize counties into five vulnerability levels: minor, moderate, major, severe, and extreme (See Methods for further detail).

For the power system vulnerability values, the probability density curve exhibits a prominent peak near the lower values, with a long tail extending to the right (Fig 2c). This pattern reveals that most values are clustered around the lower end, while a few higher values stretch the distribution, resulting in a right-skewed curve. The peak indicates that the most likely values fall within the range of [5, 10] (Min: 1, Max: 100), suggesting that most counties experienced relatively low power system vulnerability. For values greater than 30, the distribution tail thins significantly but extends to 100, indicating the presence of some counties with extremely high power system vulnerability values.

The second index is the power system vulnerability scores and distinct spatial patterns emerge in its spatial map (Fig 2a). The US West Coast (particularly California), the East Coast (including the Northeast megalopolis and Florida), the Gulf of Mexico (mainly Texas), and the Great Lakes megalopolis exhibit the highest vulnerability scores, indicating severe power outage impacts in these areas. In addition, high scores are present in many central regions, such as Colorado, Minnesota, and Wyoming, suggesting that power system vulnerability is widespread across the US counties.

The third index involves clustering the 3022 counties into five power system vulnerability ratings. The K-means clustering algorithm effectively distinguished these five categories, where the values are well-separated (Fig 2e). A total of 2704 counties (89.48% of all counties in this study) was classified as having minor or moderate levels, indicating that the majority of counties exhibit relatively low power system vulnerability (Fig 2d). Counties classified as having major, severe, and extreme vulnerability levels totaled 253 (8.37%), 54 (1.79%), and 11 (0.36%), respectively. These counties are primarily located in California, the Northeast megalopolis, Florida, the Great Lakes megalopolis, and Texas (Fig 2b). The 11 extreme-level counties are Los Angeles County, CA (100.00); Miami-Dade County, FL (99.97); Waynesboro City, VA (99.93); Niobrara County, WY (99.9); Buena Vista City, VA (99.87); Wayne County, MI (99.83); Harris County, TX (99.80); Broward County, FL (99.77); San Bernardino County, CA (99.74); Orange County, CA (99.70); and Riverside County, CA (99.67). Most of these counties are located in metropolitan areas and are typically susceptible to extreme weather-induced events. Beyond the well-documented stresses of natural hazards, social vulnerability, and segregation [28-30], this study demonstrates that the population in these counties also endure a significant risk of power outages.

To provide deeper insights, we aggregated the county-level ratings to the state level (Fig 3). Six states were identified as having counites of extreme power system vulnerability level: California (number of counties=4), Florida (n=2), Virginia (n=2), Texas (n=1), Michigan (n=1), and Wyoming (n=1). Notably, 45 states (91.84% of 49 states included in this study) contain counties classified as having major, severe, or extreme level of power system vulnerability, and 22 states (44.9%) include counties with severe or extreme level, indicating widespread electricity disruption risk across the US. We also calculated the proportion of counties in the severe and extreme levels relative to the total number of counties at the state level. The top three states are California (24.56%), New Jersey (23.81%), and Florida (14.93%), suggesting that these states not only face extensive but also a greater extent of power system vulnerability compared to others.

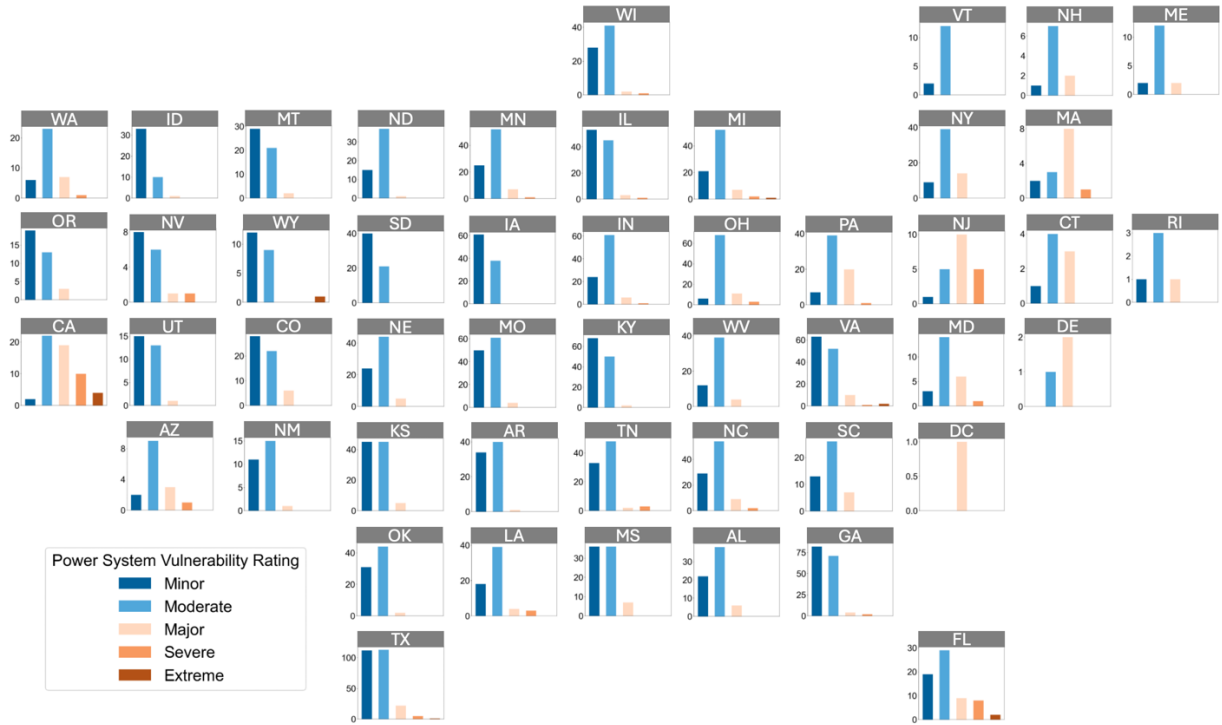

**Fig 3. Spatial distribution of the power outage vulnerability ratings at the state level.** We clustered the county-level power system vulnerability values into 5 categories (minor, moderate, major, severe, and extreme) using K-means. The bar plots show the number of counties in each category for every state, with state abbreviations displayed at the top of each subplot. A total of 48 states and Washington, D.C. were mapped according to their approximate geographic locations.

**Annual spatiotemporal patterns in power system vulnerability**

The power outage dataset covers the period from 2014 to 2023, which not only enabled the establishment of the PSVI spanning the entire decade but also allowed for the calculation of annual power system vulnerability values, scores, and ratings, facilitating year-over-year spatiotemporal comparisons and trend analysis. The annual spatial distribution of the power system vulnerability scores is available in Supplementary Fig 5.

Spatially, we examined counties that repetitively experienced high power system vulnerability over the years. We set thresholds for the number of years a county faced major, severe, or extreme vulnerability levels ($\geq 2$, $\geq 4$, $\geq 6$, $\geq 8$, and $=10$). Hotspots were then defined as counties where the cumulative number of years exceeded these thresholds throughout the study period (Fig 4a). From 2014 to 2023, a total of 333 counties (11.02% of all counties in this study) were identified as hotspots that consistently experience high levels of power system vulnerability. These counties are primarily located in regions such as the Northeast area, Florida, Texas, California, and Washington. Notably, 22 counties (0.73%) experienced persistent high vulnerability for 10 years, while 58 counties (1.92%) experienced this for 8+ years, 109 counties (3.61%) for 6+ years, and 182 counties (6.02%) for 4+ years. In California, 38 counties (66.67%, 57 counties included in this study) faced high power system vulnerability levels for 2+ years, with 6 counties (Los Angeles County, Riverside County, Sacramento County, San Bernardino County, San Diego County, and Ventura County) classified as major, severe, or extreme for 10 years. Similarly, Florida had 31 counties (46.27%, 67 counties included in this study) with high vulnerability for 2+ years, and 6 counties (Broward County, Duval County, Hillsborough County, Miami-Dade County, Palm Beach County, and Pinellas County) consistently rated as major, severe, or extreme throughout the decade.

To analyze temporal trends, we first plotted the boxplot distributions of annual power system vulnerability values (Fig 4b). Over the years, the interquartile ranges (Q1-Q3) of the boxplots consistently shift to the right, indicating a steady increase in power system vulnerability. Notably, the average annual increase rate has been significantly higher since 2019 compared to the previous years (2014-2019: 9.86% vs. 2019-2023: 18.84%), with the most pronounced increase occurring between 2022 and 2023 (2022-2023: 37.80%).

We also aggregated the county-level power system vulnerability values to the state level, calculating the annual average for each state (Fig 4c). The overall trend of the state-level averages also exhibits a consistent increase, mirroring the annual county-level pattern in Fig 4b. However, certain states deviate from this pattern, which show spikes in some specific years. For example, California experienced persistently high vulnerability between 2017 and 2019. Similarly, Connecticut saw a sharp increase in 2018, while New Jersey recorded high values in 2021 and 2022, and Delaware, Maine, and New Hampshire displayed spikes in 2023. These spikes may correspond to extreme weather-induced events: California experienced record-breaking wildfires from 2017 to 2019 [31], while Connecticut, New Jersey, Delaware, Maine, and New Hampshire were affected by winter storms, tropical cyclones, or extreme rainfall during the years [32-34]. The impact of extreme weather-induced events on power system vulnerability is profound. Florida experienced high vulnerability values in 2016, 2020, and 2022, which align with Hurricane Matthew in 2016, Hurricane Sally in 2020, and Hurricane Ian in 2022 [35]. Similarly, Texas saw vulnerability spikes in 2021 and 2023, corresponding to the deadly winter storm in 2021 and the record-breaking heatwave in 2023 [5,36].

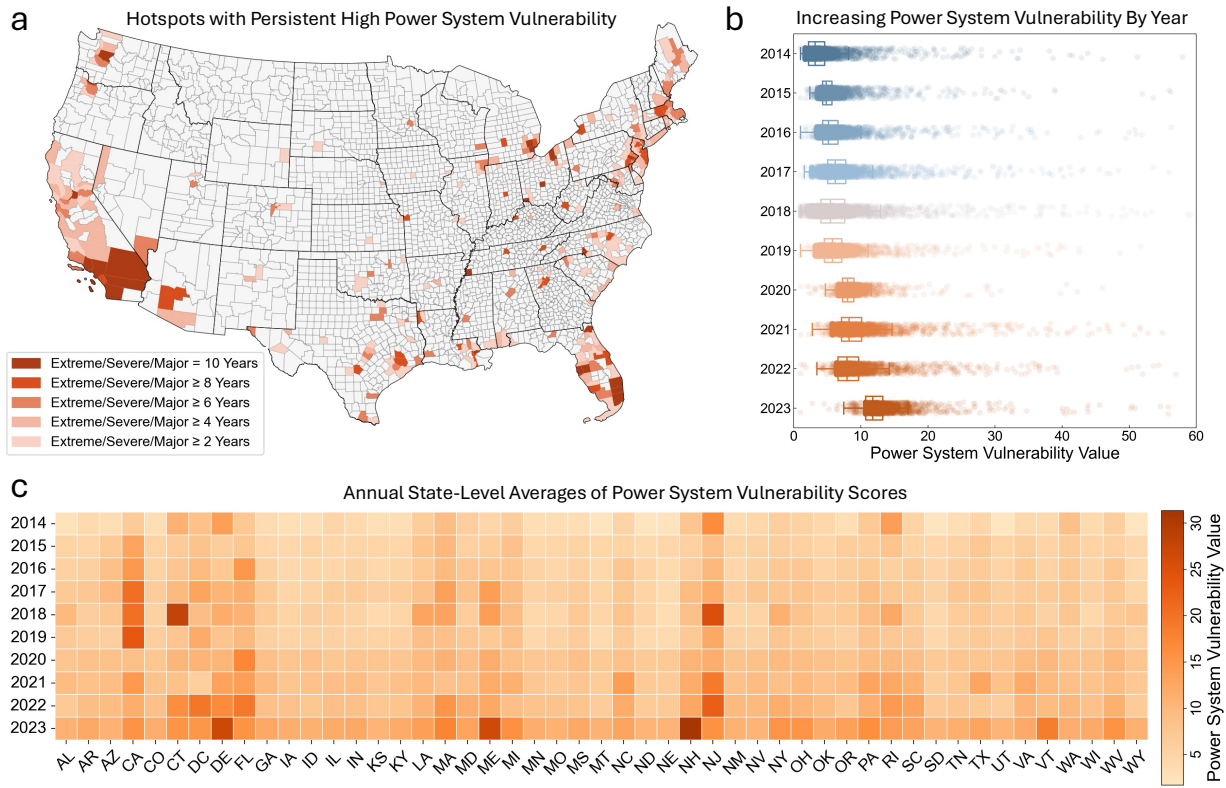

**Fig 4. Spatiotemporal distribution of the annual power system vulnerability index. a.** Spatial distribution of counties with persistent high power system vulnerability. We set thresholds for the number of years a county faced major, severe, or extreme levels (≥2, ≥4, ≥6, ≥8, and =10). Hotspots were defined as counties where the cumulative number of years exceeded these thresholds between 2014 and 2023; **b.** Boxplot of annual distribution of the power system vulnerability values. The Kruskal-Wallis H test confirmed the significant differences among the ten years ($p<0.001$); **c.** Heatmap of the annual average for the power system vulnerability values at the state level. The X-axis shows the abbreviations of the 48 US states and Washington, D.C.

**Analysis of disparities in power system vulnerability**

To investigate the spatial heterogeneity of power system vulnerability, we first applied the cumulative probability density function to analyze the relationship between power system vulnerability and urbanicity. Counties were classified as either urban or rural based on the 2013 National Center for Health Statistics (NCHS) Urban-Rural Classification Scheme (Supplementary Fig 6) [37]. In Fig 5a, the distributions between urban counties (n=1776) and rural counties (n=1246) are basically consistent, with the majority of the counties (>90%) exhibiting low power system vulnerability values (<20). Subsequently, the slopes of the distributions start to increase exponentially, indicating drastically greater power system vulnerability values (>20) for a smaller proportion of counties (<10%). The long-tail distributions show the heterogeneity of power system vulnerability within each group. When comparing the two distributions, the curve of the urban counties consistently sits above that of the rural counties. This result reveals not only a disproportional distribution of power system vulnerability among all the counties but also that the urban counties have greater power system vulnerability compared with rural counties.

We further examined the extent to which the form and structure characteristics of urban and rural areas shape the heterogeneity in power system vulnerability. Our analysis involved three urban/rural form and structure dimensions: development density (DD), centrality & segregation (CS), and economic activity (EA) [28]. DD encompasses factors such as population density, POI (point of interest) density, and road density; CS includes urban centrality index, minority segregation, and income segregation; and EA considers GDP and human mobility index (see Methods for further detail). In rural counties, only DD exhibited a significantly positive correlation with PSVI (0.20***), whereas, in urban counties, both DD (0.59***) and EA (0.25***) showed significantly positive correlations (Fig 5b). The positive correlation of DD indicates that higher

DD in both rural and urban areas contribute to increased power system vulnerability. The effect of DD was greater in urban counties compared to rural counties (0.20 vs. 0.59), reflecting the heightened power system vulnerability in cities due to denser population, facilities, and roads. For example, compared to rural areas, dense street trees and lights in cities are more easily blown down by strong winds, destroying power grids and causing power supply disruptions. Moreover, urban counties show a strong correlation between EA and PSVI. Economic activities are heavily dependent on electricity, which may be significantly disrupted by outages.

To assess the impact of different power grid connectivity scenarios on power system vulnerability, we categorized counties based on regional electricity transmission data. The US has seven Regional Transmission Organizations (RTOs) that consolidate high-voltage transmission assets to enhance efficiency across a large network (Supplementary Fig 7) [38]. As cases are possible that counties belong to multiple RTOs, we labeled these as boundary counties. Fig 5c presents a boxplot distribution of the power system vulnerability values across the different RTOs. CAISO (serves California) exhibits the highest power system vulnerability, followed by ISO-NE (serves New England regions), and NYISO (serves New York state). ERCOT (serves Texas), which was severely attacked in the 2021 Winter Storm [5], shows a relatively low vulnerability in this analysis, probably because our data spans a decade and accounts for more than just isolated extreme events. Interestingly, boundary counties also show moderately high power system vulnerability, possibly due to their locations at the intersection of multiple transmission lines, which may contribute to instability in power supply. This analysis reveals the spatial heterogeneity of power system vulnerability in power grid connectivity and highlights that areas with interconnected grids are sensitive to cascading failures during disruptions.

With the rapid expansion of renewable energy sources like solar and wind, it is essential to examine how the development of these energy technologies impacts power system vulnerability. We collected the decadal average of net generation percentages for solar and wind at the state level [39] and compared them with the power system vulnerability values through Ordinary Least Squares (OLS) regression. In Fig 5d, solar energy demonstrated a significant positive correlation with power system vulnerability ($p<0.001$, $R^2=0.3666$), indicating that states with higher solar generation tend to experience greater vulnerability of power systems. For example, California heavily relies on solar energy [40], and also shows higher power system vulnerability. However, wind energy shows an inverse relationship with power system vulnerability, with states that generate more wind energy experiencing less vulnerability. Such relationship is less interpretable than that with solar energy ($p<0.01$, $R^2=0.1669$). Wind resources are primarily concentrated in the central regions and coastal areas of the US, while coastal areas have seen slower development of wind generation [41]. For example, Florida has yet to pass legislation to permit wind generation [42]. Louisiana, despite its abundant offshore wind resources, only began developing wind generation in 2023 [43]. California imports a significant portion of wind energy from other states, which does not count as in-state wind generation [44]. Consequently, these states with a higher power system vulnerability tend to have a lower proportion of wind energy, while the central regions with abundant wind generation exhibit lower power system vulnerability. This contrast contributes to the differing impacts of wind energy compared to solar energy.

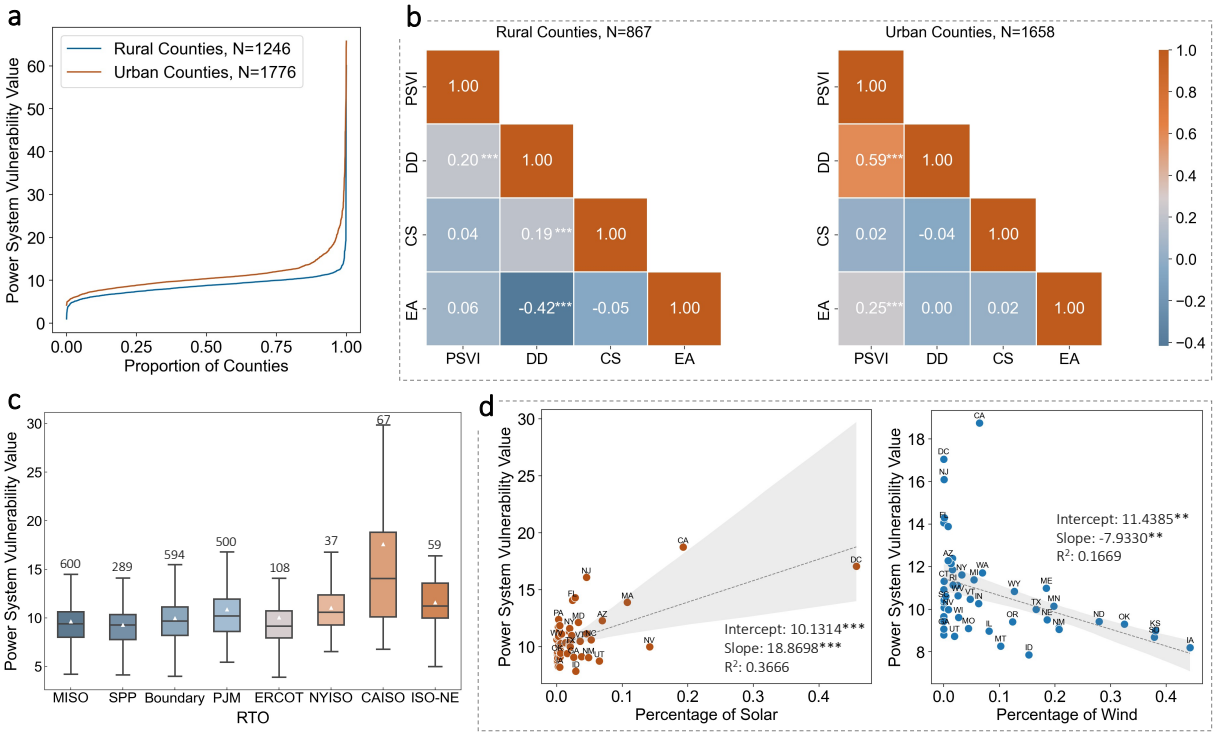

**Fig 5. Disparity of the power system vulnerability index. a.** Cumulative probability density curve of the power system vulnerability values between rural and urban counties. The one-way ANOVA test confirmed the significant differences between the two groups ($p<0.001$); **b.** Heatmaps of Pearson correlation between the urban/rural form and structure dimensions and PSVI. Each number represents a correlation coefficient; **c.** Boxplot of the power system vulnerability values among different RTOs. Counties belong to multiple RTOs were labeled as boundary. The number above each box refers to the number of counties and the white arrow represents the average of the power system vulnerability values for that category. The Kruskal-Wallis H test confirmed the significant differences among the groups ($p<0.001$); **d.** OLS regression plots between the power system vulnerability values and percentage of solar/wind energy used for generation. The shaded area denotes a 95% confidence interval. The "***" represents $p<0.001$ and "**" represents $p<0.01$.

## Discussion

Despite considerable evidence regarding the vulnerability of the US power infrastructure [2-5], a systematic and national-level assessment of the spatiotemporal patterns of power system vulnerability has still been lacking. To address this, we retrieved ~179 million power outage records with 15-minute intervals between 2014 and 2023 to conduct a county-level evaluation of power system vulnerability. We proposed a power system vulnerability assessment framework based on three dimensions: intensity, frequency, and duration. Using interpretable machine learning models of XGBoost and SHAP, we generated the PSVI for 3022 counties, covering 96.15% of the area. Our results reveal distinct spatiotemporal patterns of power outage vulnerability, which have consistently increased over the past decade. Urban counties, CAISO, and states with high solar generation showed significantly higher vulnerability levels. The findings provide critical empirical evidence of the pervasive and widespread nature of power outages across the US and offer valuable insights for stakeholders to guide the development of policies and programs aimed at strengthening the resilience of the US power infrastructure.

Few studies have examined power system vulnerability over long-term periods due to the lack of large-scale and high-resolution outage data. Prior case-based research focused on quantifying the short-term impacts of power outage exposure caused by extreme weather events, such as wildfires, hurricanes, or heat waves [3,5,10]. However, concentrating solely on isolated events may obscure longer-term trends that are critical for informed policy decisions in the context of climate change. In addition, existing studies often limits the geographic scope to specific regions or states [12-14], lacking a nationwide sub-state level assessment capable of capturing the heterogeneity in outages across various geospatial contexts. Furthermore, a data-driven characterization of power system vulnerability requires generalizable metrics and systematical assessment models, both of which are underdeveloped in the current literature.

This study addresses the existing gaps by leveraging a decadal outage data from EAGLE-I$^{TM}$ [21]. The dataset covers all outage records at 15-minute intervals from 2014 to 2023 in 3022 US counties, capturing both large-scale outage events caused by extreme weather events and frequent but localized outage events resulting from system aging, physical attacks, routine maintenance etc. The proposed power system vulnerability assessment framework, based on the environmental hazards exposure theory [22-24], integrates multiple dimensions of exposure: frequency, duration, and intensity. By integrating commonly used features from previous studies with those specifically designed for large-scale outage events, we developed a systematic and comprehensive set of 14 features to characterize the extent of power outages. To enhance differentiation, we applied multiple machine learning models (XGBoost, SHAP, and K-means) to generate three distinct PSVIs: values, scores and ratings. To the best of our knowledge, the PSVI is the first nationwide assessment of power system vulnerability. The index aims to provide users, ranging from local, regional, state, and federal planners to emergency managers and other decision-makers, with a clear understanding of the power outage risk in their communities. With publicly available data and methods, our index is both practical and transparent, ensuring reproducibility and usability by researchers and practitioners.

We observed a consistent increase of power system vulnerability in US counties over the past decade. Previous studies have suggested that the rise in severe weather events due to climate change will likely lead to more frequent power outages in the US [5,11]. Our multidimensional analysis of power outages confirms that the PSVI at the county level has steadily risen from 2014 to 2023. Specifically, we found a significant increase in the annual PSVI in the five years following 2019 compared to the preceding five years, with the most pronounced spike occurring between 2022 and 2023. At the state level, the higher power system vulnerability corresponds with increasing frequency of natural hazards. It is important to note that although climate change-induced nature hazards drive recent large-scale outages, factors such as the aging grid and rising energy demand contribute to smaller-scale but frequent outages [6]. While our study identifies the existence of relationship between natural hazards and outage events, examining the interaction between the factors (i.e., climate change, aging grids, energy demand) and power outages falls beyond the scope of this paper.

A robust literature has established that power outages disproportionately affect certain regions [2,13,14]. In this study, the distribution of the PSVI follows a long-tail pattern, where the majority of counties exhibits low vulnerability while a smaller proportion experiences extremely high vulnerability. These high-vulnerability areas are concentrated on the West Coast (particularly California and Washington), the East Coast (notably Florida, the Northeast metropolitan area, and the New England areas), the Great Lakes megalopolis (mainly Chicago-Detroit area), as well as the Texas Gulf Coast area. The PSVI offers greater precision and scope than prior studies [2], enabling us to capture longstanding spatial heterogeneity on a national scale. Our findings indicate that urban counties generally have higher power system vulnerability than rural counties. In urban environments, higher vulnerability to outages can be attributed to dense infrastructure development, large-scale electricity demand due to population concentration, and the proximity of power grids to critical facilities. Such pattern increases the likelihood of cascading failures from equipment malfunctions, accidents, or construction activities.

In addition, our study reveals significant variation in power system vulnerability across different RTOs and states with diverse energy structures. We found that CAISO exhibited markedly higher power system vulnerability compared to the other six US RTOs. The resilience of the grid within CAISO has been extensively studied in the prior work [45,46]. Our analysis also highlights the high vulnerability of counties located along the borders of regional transmission grids, suggesting that these "gray areas" warrant particular attention. Furthermore, our study analyzed the ongoing energy transition by examining the relationship between power system vulnerability and the proportion of renewable energy generation at the state level. We observed that states with higher solar generation experienced greater power system vulnerability. Given the essential role of energy in daily life, such correlations in power system vulnerability raise significant energy inequality concerns. By exploring these disparities, our study contributes to the growing literature on environmental justice, highlighting an issue that has not yet been sufficiently addressed—namely, the inequitable distribution of power outages and restoration efforts [47-49].

The findings offer multiple important contributions and implications to the interdisciplinary fields of energy systems, infrastructure engineering, and disaster resilience. First, by introducing PSVI, we provide the first comprehensive, data-driven, and quantitative tools specifically designed to evaluate power system vulnerability across all US counties. This outcome addresses a critical gap in existing research, where numerous spatial indexes capture social and physical vulnerabilities, but none adequately represents the vulnerability of power systems themselves. Our approach has important implications for preparedness strategies by enabling emergency managers and utilities to leverage the index before impending extreme events, identifying high-risk counties with unprecedented precision, and allowing for targeted and proactive allocation of restoration crews and resources. Furthermore, the integration of the PSVI into hazard mitigation and resilience planning would significantly enhance these processes. Second, our quantification of power systems vulnerability based on historical outages offers a deeper and more nuanced characterization of spatiotemporal variations across counties, unveiling patterns previously less understood. The study not only reveals hotspots of power outages but also identifies areas where vulnerability has consistently increased over the past decade. Such detailed spatiotemporal insights are crucial for informing regional power system resilience plans and shaping national policies aimed at mitigating risk, offering a level of granularity and temporal depth that was previously unattainable. Third, the findings illuminate complex variations in power systems' vulnerability across urban-rural gradients and pinpoint specific structural and developmental features that influence vulnerability levels. For instance, identifying development density as a key determinant of power system vulnerability underscores the pressing need for tailored resilience measures in rapidly growing urban areas with dense development. This revelation highlights the importance of the relationship between urban planning and infrastructure vulnerability and demonstrates how urban development contributes to power system vulnerability. Finally, the study breaks new ground by quantifying the potential impact of renewable energy sources on power system vulnerability. By highlighting this previously unexplored relationship, we open new areas of inquiry regarding the role of renewable energy in shaping vulnerability assessments and resilience strategies, paving the way for a more holistic understanding of energy insecurity and its intersections with climate change and disaster resilience. Collectively, these contributions and implications offer data-driven insights and practical metrics that empower power infrastructure owners, operators, emergency managers, and public officials to effectively address the escalating vulnerability of power infrastructure across the US.

There were also limitations in this study, which could be addressed in the future. First, the county-level assessment does not capture finer sub-county-level details such as increasing development and population influx over the years. While we identified spatial heterogeneity in power system vulnerability, the trends might differ with more granular data. If higher resolution data becomes available, future studies could benefit from examining sociodemographic characteristics and other metrics at finer spatial resolutions to better understand inequalities in power outage extent, especially for marginalized communities. Second, the power grid characteristics across different counties were not fully considered in this study. The power grids comprise complex networks with interdependent relationships between transmission and distribution grids of various sizes. While our study considered power grid connectivity as a distinguishing factor, if grid characteristics data is available, future research could incorporate more detailed data on transmission and distribution grids to better differentiate spatial variations and spillover effects in power system vulnerability.

# Methods

This study follows the processing procedure shown in Fig 6. First, we developed 14 power system vulnerability features from three dimensions (frequency, intensity, and duration), capturing multifaceted outage characteristics. Next, we applied interpretable machine learning models (XGBoost and SHAP) to determine the relative importance of features. Based on the feature importance, we assigned corresponding weights to features. Then, the PSVI (value, score, and rating) was calculated as a weighted sum of features. Finally, we conducted disparity analysis of power system vulnerability regarding factors including urban/rural form and structure, power grid connectivity, and electricity generation by source.

**Power outage data**

This study utilized a large-scale and high-resolution power outage dataset to calculate power outage-related features across US counties. Power outage data was collected through Environment for Analysis of Geo-Located Energy Information (EAGLE-I$^{TM}$) by Oak Ridge National Laboratory (ORNL) [21]. EAGLE-I$^{TM}$ compiled electricity service disruption records from individual electrical utilities at a 15-minute interval from 2014 through 2023. On average the dataset covers ~90% of utility customers nationwide, making it the most comprehensive outage information ever compiled in the US [21]. Considering that data for Alaska, Hawaii, Puerto Rico, Guam, the US Virgin Island, and American Samoa are incomplete in certain years, we limited the geographical range of our study within the contiguous United States. The data covers 3022 counties, accounting for 96.15% of the area.

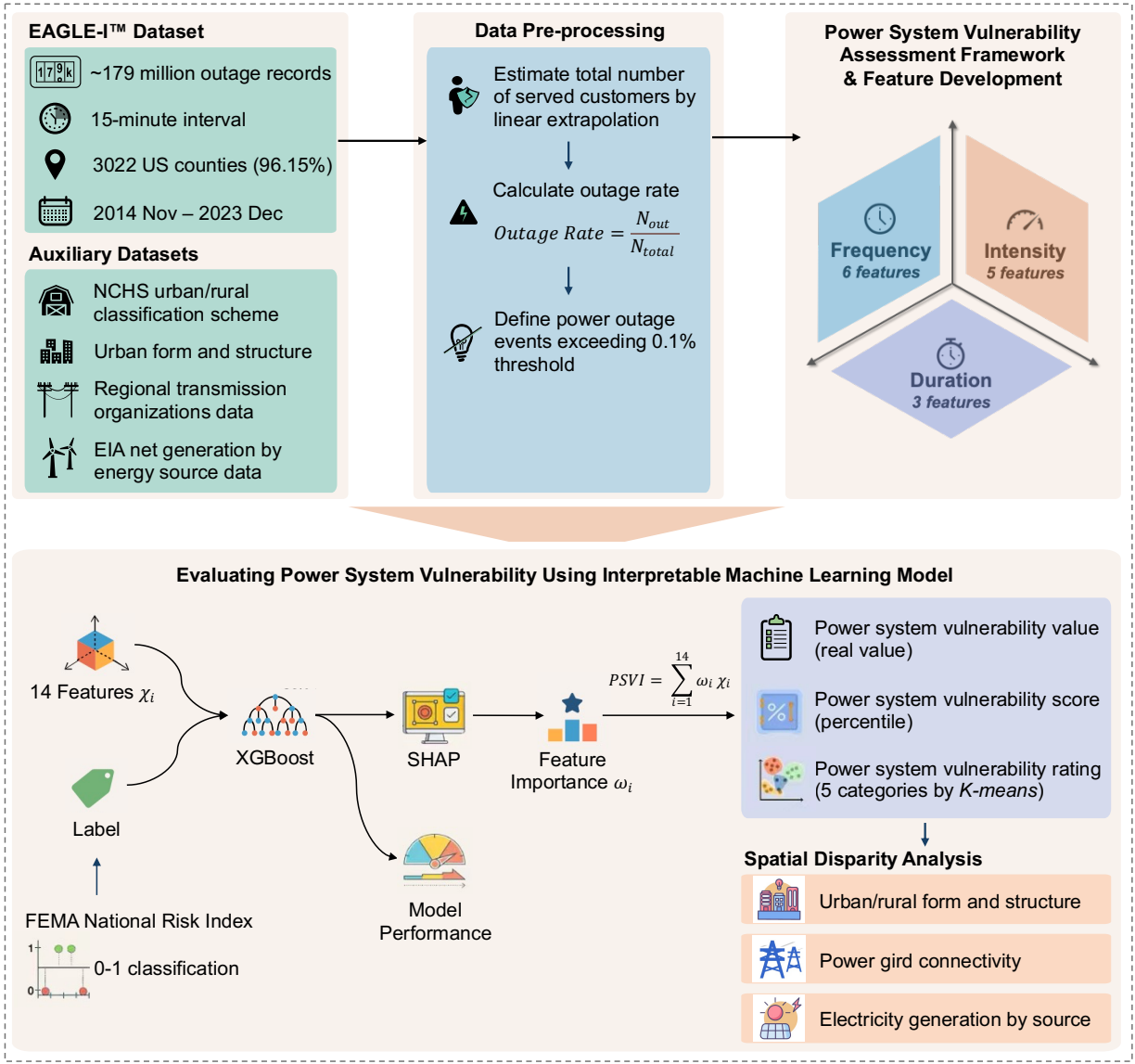

**Fig 6. Workflow for establishing power system vulnerability index across US counties using large-scale power outage data and interpretable machine learning models**

**Power outage features development**

We processed the outage records following the procedure described in our previous study [50] to calculate power outage features. We defined power outage events as the continuous time period during which power outage rate exceeds 0.1%. Setting the 0.1% threshold helps to screen out the non-valid outage records due to incidental factors, which is a practice applied by research such as Do, McBrien et al. (2023) [2]. Power outage rate is calculated as the proportion of customers without power compared to total number of customers in a county (Equation 1). The ORNL provided the estimated number of total customers for certain temporal and geographical range, and we utilized linear extrapolation to extend the data to cover all the 3022 counties from 2014 through 2023. Over the decade, a total of 3,022,915 power outage events was identified.

$$Power\ Outage\ Rate = \frac{N_{out}}{N_{total}} \quad (1)$$

where, $N_{out}$ refers to the number of customers experiencing outages; $N_{total}$ represents the total number of customers in a county.

We proposed a systematic and comprehensive power system vulnerability assessment framework, which quantified outages through three dimensions: frequency, duration and intensity. Under each dimension, multiple features were developed to capture the characteristics of power outages (Fig 1). To account for the impacts of large-scale outage events, which affect daily life more significantly than frequent but localized events, we set thresholds based on average outage rates (5%) and average duration (12 hours) to define features for large-scale outage events. All the features were calculated at the county level. For the decade PSVI, the features were aggregated and calculated over the ten years, while for the annual PSVI, the features were calculated on a yearly basis. The statistical summary of these features is available in Table 1, and spatial distribution maps over the ten years are provided in Supplementary Fig 4.

- **Intensity.** This dimension incorporates five features: (1) average outage rate—calculated as the average outage rate among all power outage events; (2) cumulative number of customers affected—counted as the total number of customers affected in all power outage events; (3)

peak number of customers affected—counted as the maximum number of customers affected in a single outage event. This feature captures the historical peak outage intensity of the counties; (4) average increase/decrease rate. To calculate this feature, we aggregated raw outage records to a monthly level and computed the percentage changes between consecutive months. The average increase/decrease rate is calculated as the average of percentage changes, reflecting how drastically the outage records were changing over time; (5) average outage rate exceeding 12 hours—calculated as the average outage rate among power outage events whose duration exceeded 12 hours.

- **Frequency.** This dimension includes six features: (1) number of events—counted as the total number of power outage events; (2) average inter-event time—calculated as the average time interval between consecutive outage events; (3) number of events affecting >5% customers—counted as the total number of outage events which affected more than 5% of the served customers; (4) average inter-event time affecting >5% customers—measured as the average time interval between outage events which affected more than 5% of the served customers; (5) number of events exceeding 12 hours—counted as the total number of events which lasted more than 12 hours; (6) average inter-event time exceeding 12 hours—measured as the average inter-event time between outage events whose duration exceeded 12 hours.
- **Duration.** This dimension contains three features: (1) average duration—calculated as the average duration of outage events; (2) average duration per customer experienced—calculate by dividing the total outage duration by the total number of customers. This is a normalized outage duration feature regarding the scale of served customers; (3) average duration affecting >5% customers—calculated as the average duration of outage events that affected more than 5% of served customers.

**Table 1. Statistical summary of the power system vulnerability features.** The summary represents the county-level averages over the study period (2014–2023), based on data from 3022 US counties.

| Features | Unit | Mean | Max | Min | Median |
|---|---|---|---|---|---|
| Number of events | / | 1002.3 | 5772.0 | 1.0 | 891.0 |
| Average outage rate | % | 1.5 | 100.0 | 0.1 | 1.0 |
| Average duration | days | 7.3 | 81.6 | 0.6 | 6.6 |
| Average inter-event time | days | 7.1 | 273.5 | 0.0 | 3.4 |
| Cumulative number of customers affected | / | 540915 | 32050993 | 2 | 191557 |
| Peak number of customers affected | % | 52.8 | 100.0 | 0.1 | 49.6 |
| Average increase/decrease rate | % | 501.5 | 10121.5 | -17.0 | 254.7 |
| Average duration per customer experienced | hours | 0.4 | 52.6 | 0.0 | 0.1 |
| Number of events affecting >5% customers | / | 42.0 | 1591.0 | 0.0 | 28.0 |
| Average duration affecting >5% customers | days | 0.4 | 15.9 | 0.0 | 0.23 |
| Average inter-event time affecting >5% customers | days | 228.3 | 10075.7 | 0.0 | 119.9 |
| Number of events exceeding 12 hours | / | 29.2 | 394.0 | 0.0 | 19.0 |
| Average outage rate exceeding 12 hours | % | 2.2 | 18.3 | 0.0 | 1.7 |
| Average inter-event time exceeding 12 hours | days | 136.8 | 3062.1 | 0.0 | 88.7 |

**Power system vulnerability index construction**

We constructed the PSVI as the weighted sum of the power system vulnerability features with the Equation 2.

$$PSVI = \sum_{i=1}^{14} w_i x_i \qquad (2)$$

where, $x_i$ represents the power system vulnerability feature; $w_i$ denotes the weight of feature $x_i$.

We weighted the features according to their relative importance in contributing to the impact of power outages. In other words, the more a feature contributes to the impacts, the greater weight should be assigned to it, as it significantly shapes the power systems' vulnerability. We used the NRI [26] as a proxy for power outage impacts, since the indicator is a comprehensive assessment of overall risk for US counties by integrating multiple hazards and vulnerability factors into a singular metric. By training the XGBoost model to predict the NRI based on our features, we effectively gauged the extent to which power system vulnerability features explain variations in the county-level risk. Accordingly, we used the relative importance of the features retrieved from SHAP model as weights for computing the PSVI. This approach to determining the weights of features instead of relying upon subjective weights provides a more reliable estimation of the weights in calculating the PSVI [20,51].

Before performing the XGBoost model, we preprocessed the data through the following pipelines: (1) data labelling. The NRI rates all counties into five risk categories: very low, relatively low, relatively moderate, relatively high, and very high. We labelled counties in "very low" category as label 0 (n=1419), and counties in other categories as label 1 (n=1603); (2) feature normalization. We rescaled the power system vulnerability features to [0, 1] using min-max normalization for consistency; (3) multicollinearity check. Features with high multicollinearity share similar information about the target variable, which could cause redundancy and complicate interpretation [52]. Thus, we calculated the variance inflation factor (VIF) to diagnose multicollinearity issues. VIF greater than 10 indicates high multicollinearity in the dataset [53]. In this study, the VIF of power system vulnerability features are all smaller than 10, indicating that multicollinearity is not a significant issue (Supplementary Table 1).

For model training and validation, we split the data into 80/20 ratio, with 80% of data used for training and the remaining 20% for testing. To improve the model performance, we applied Synthetic Minority Oversampling Technique (SMOTE) to mitigate the impact of category imbalance by over-sampling the minority category and under-sampling the majority category [54]. Also, we applied random search and performed a 10-fold cross-validation to tune the hyperparameters. The tuned best hyperparameters for the XGBoost model is listed in Supplementary Table 2. We compared the nine widely used classification models, such as random forest, support vector machine, and AdaBoost, and found that XGBoost achieved the best F1-score (Supplementary Table 3). Hence, we selected XGBoost as our primary model for binary classification.

To interpret how the power system vulnerability features contribute to the impacts of power outages, we adopted SHAP model. SHAP value of each feature denotes both the magnitude and direction of contributions towards the machine learning output [27]. In this study, SHAP values were calculated as a measure of feature importance and rescaled to [0, 1] (Fig 2g). We used the following equation to convert SHAP values into feature weights to make sure the sum of the weights equals 1:

$$w_i = \frac{SHAP_i}{\sum_{i=1}^{14} SHAP_i} \tag{3}$$

where, $SHAP_i$ is the SHAP value of the feature $x_i$, and $w_i$ is the weight of the feature $x_i$.

Finally, we calculated the power system vulnerability values at the county level using Equation 2. The value provides an absolute measure of power system vulnerability for each county. To enable comparison across counties, we converted these values into percentiles, referred to as the power system vulnerability scores (Fig 2a). Additionally, we established a five-category qualitative rating system (minor, moderate, major, severe, and extreme) using the K-means clustering algorithm (silhouette score=0.55) [55]. The rating system allows for a more intuitive understanding of vulnerability levels, making it easier for stakeholders to interpret the severity of vulnerability in different counties (Fig 2b).

**Auxiliary data for disparity analysis**

- **Urban/rural form and structure.** We categorized the counties as either urban or rural according to the 2013 NCHS Urban-Rural classification scheme [37]. A six-level urban-rural classification scheme for US counties and county-equivalent entities is developed by NCHS. We labeled 1776 counties as urban since they fall into the three most urban categories: large central metropolitan, large fringe metropolitan, and medium metropolitan. A total of 1246 counties were labelled as rural since they are in the three least urban categories: small metropolitan, micropolitan, and noncore (Supplementary Fig 6). Urban/rural form and structure refer to the spatial configuration and organization of regions. We examined eight form and structure features in Ma and Mostafavi (2024) [28], including population density, POI density, road density, minority segregation, income segregation, urban centrality index, GDP, and human mobility index. To reduce complexity while preserving the essential information of the features, principal component analysis was performed, and three principal components were extracted. The first principal component is named as DD, representing the level of urbanization and built environment density in certain area. The second one is CS, representing the social and economic segregation as well as the urban centralization level. The third component is EA, representing the level of economic activity and mobility. The detailed description of the features is available in Supplementary Note.

- **Power grid connectivity.** We referred to the spatial distribution of RTOs to indicate power grid connectivity. The US has seven RTOs, where utilities and other high-voltage owners pool their transmission assets to enable greater efficiency over a large network [38]. These RTOs includes California ISO (CAISO), Southwest Power Pool (SPP), Electric Reliability Council of Texas (ERCOT), Midcontinent ISO (MISO), New York ISO (NYISO), New England ISO (ISO-NE) and PJM. The spatial coverage of RTOs is shown in Supplementary Fig 7.

- **Electricity generation by source.** The US Energy Information Administration provides yearly state-level energy generation data and the share of total for each energy source [39]. Among the energy sources, we cast special attention on the wind and solar energy, as they are the most widely applied renewable electricity sources and have been reported to have impacts on the stability of power systems [56,57]. We collected relative shares of state-level electricity generation by solar and wind from 2014 through 2023 and calculated the ten-year average as the percentage of solar and wind among all kinds of energy sources.

## Data availability

The power outage data from 2014 to 2022 is publicly available at Brelsford, Tennile et al. (2024) [21], and data for 2023 can be obtained from ORNL upon request. The other datasets used in this paper are publicly accessible and cited in this paper. The 2019 TIGER/Line US County Shapefile was utilized to create the nationwide map for this study [58].

## Code availability

All analyses were conducted using Python. The code that supports the findings of this study is available from the corresponding author upon request.

## Acknowledgements

This work was supported by the National Science Foundation under Grant CMMI-1846069 (CAREER). Any opinions, findings, conclusions, or recommendations expressed in this research are those of the authors and do not necessarily reflect the view of the funding agencies.

## Competing interests

The authors declare no competing interests.

## Additional information

Supplementary material associated with this article can be found in the attached document.

# Supplementary Information for

*Establishing Nationwide Power System Vulnerability Index across US Counties Using Interpretable Machine Learning*

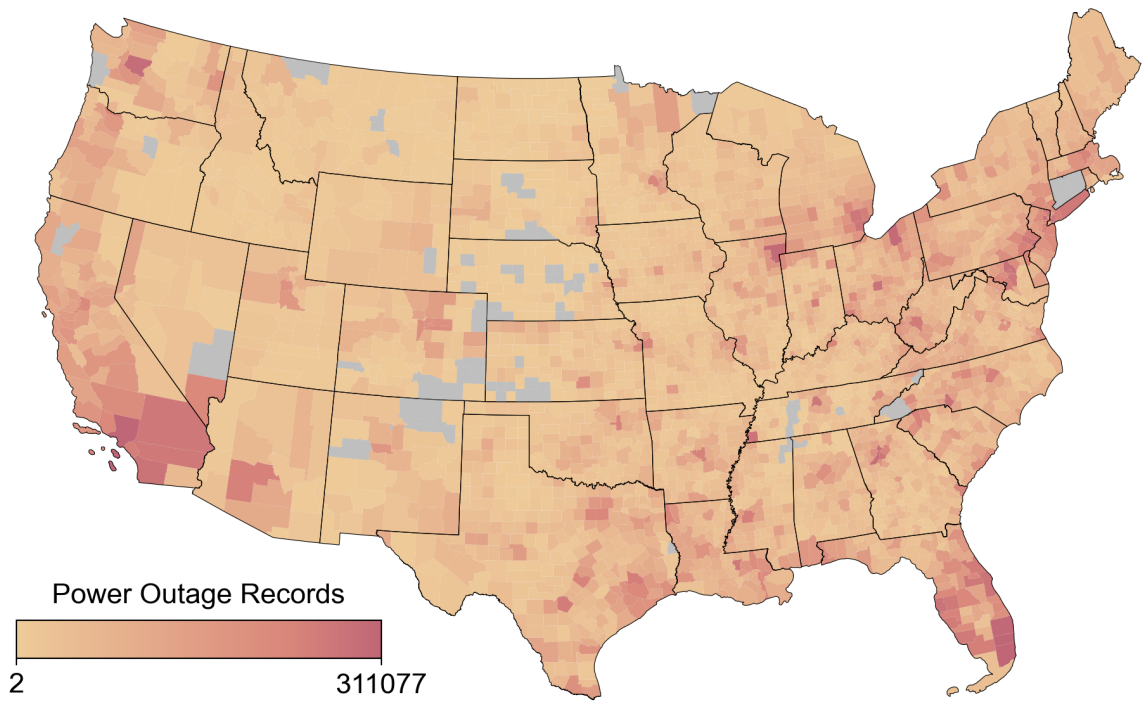

**Fig 1. Spatial distribution of the power outage records over the past decade at the county level.** From November 2014 to December 2023, we collected a total of 179,053,397 power outage records at 15-minute intervals across 3022 US counties from the Environment for Analysis of Geo-Located Energy Information (EAGLE-I[TM]). We aggregated these records at the county level over the ten years. Maps cover 3022 counties, with gray areas representing counties without data.

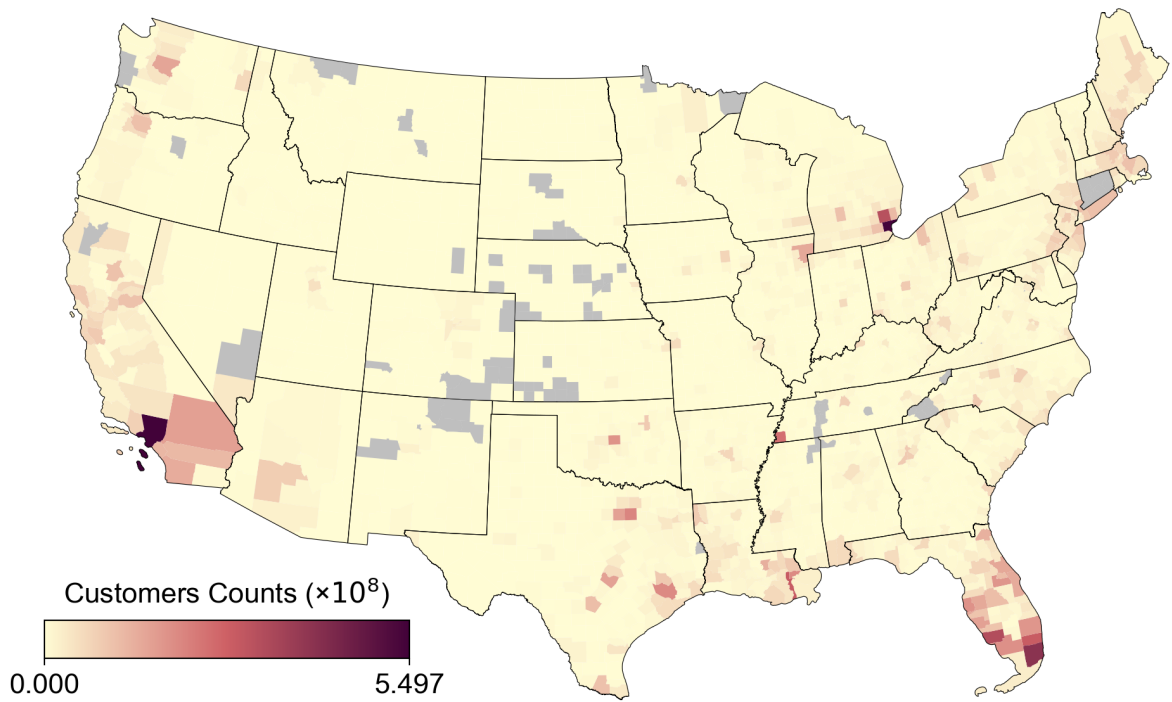

**Fig 2. Spatial distribution of the number of customers affected by power outages over the past decade at the county level.** Each power outage record contains the number of customers affected, and we aggregated these values at the county level from 2014 to 2023. The total number of customers affected across the country amounts to approximately 31.47 billion. Maps cover 3022 counties, with gray areas representing counties without data.

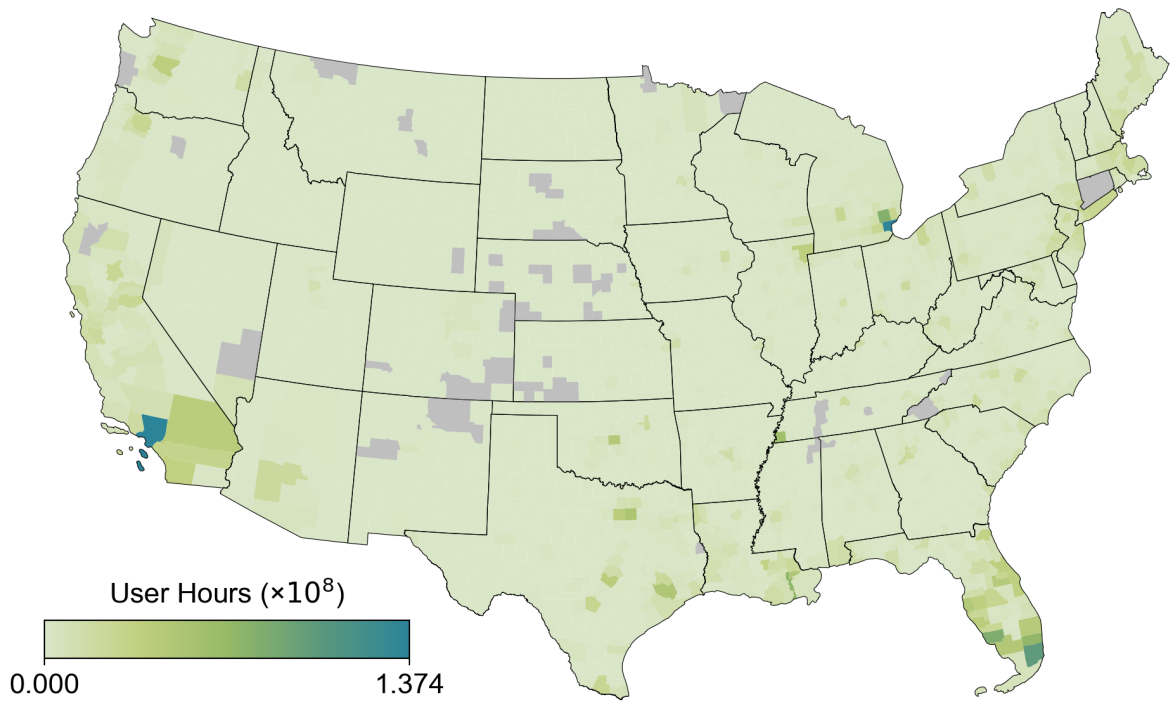

**Fig 3. Spatial distribution of the accumulative user outage time over the past decade at the county level.** The accumulative user outage time was calculated by multiplying the number of customers affected by 15 minutes (the interval) and we aggregated the results into hours at the county level from 2014 to 2023. The accumulative user outage time across the country reached 7.86 billion user-hours. Maps cover 3022 counties, with gray areas representing counties without data.

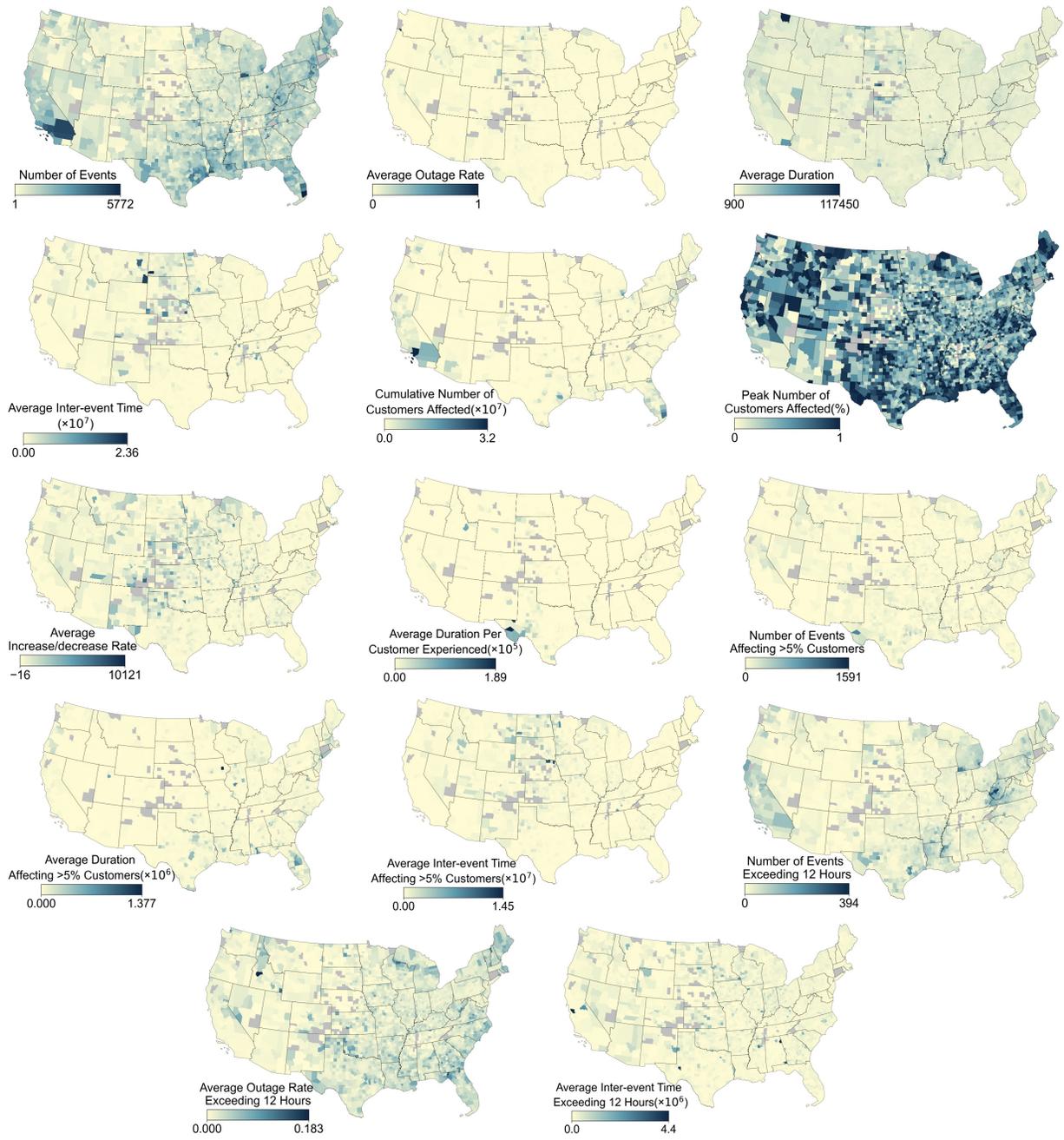

**Fig 4. Spatial distribution of power outage features over the past decade at the county level.**
Maps cover 3022 counties, with gray areas representing counties without data.

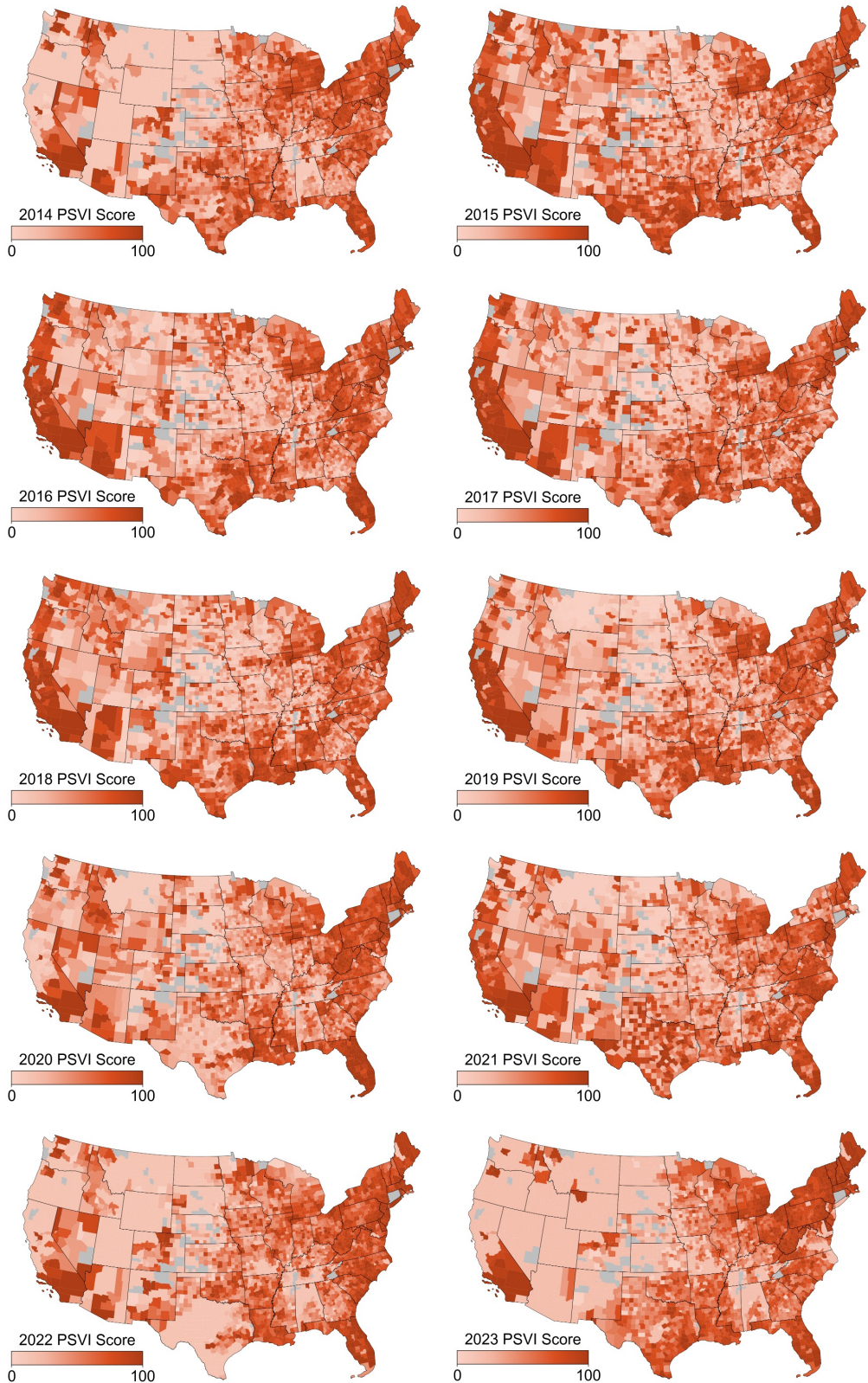

**Fig 5. Spatial distribution of power system vulnerability scores by year at the county level.**
Maps cover 3022 counties, with gray areas representing counties without data.

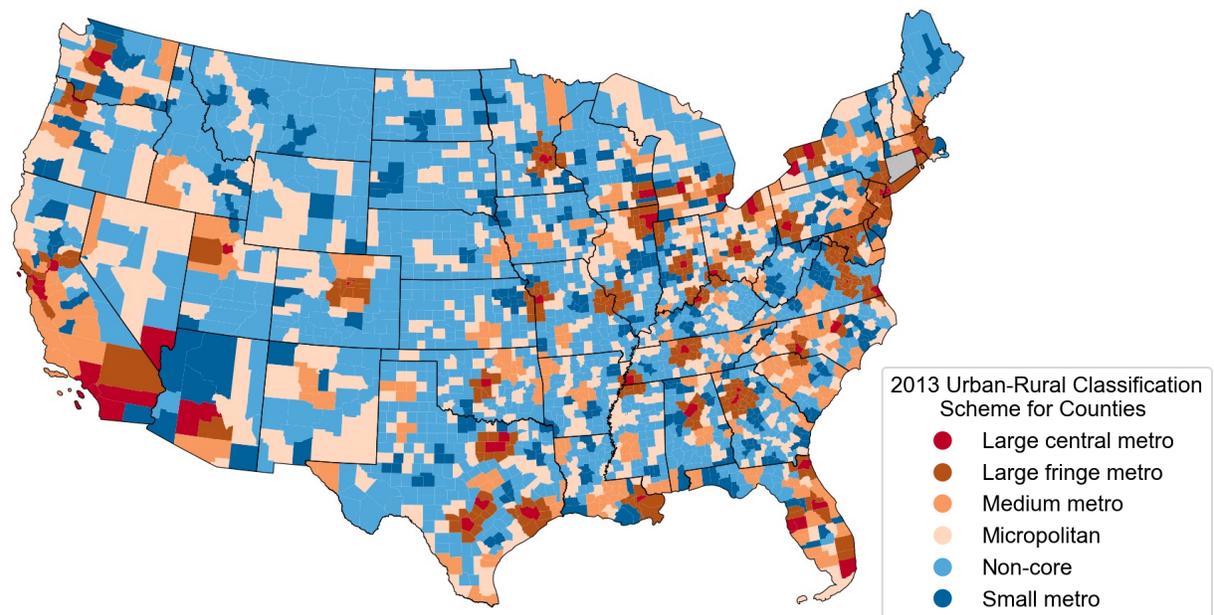

**Fig 6. Spatial distribution of urban and rural counties.** We categorized the counties as either urban or rural according to the 2013 National Center for Health Statistics (NCHS) Urban-Rural classification scheme. A six-level urban-rural classification scheme for US counties and county-equivalent entities is developed by NCHS[1]. We labeled 1776 counties as urban since they fall into the three most urban categories: large central metropolitan, large fringe metropolitan, and medium metropolitan. A total of 1246 counties were labelled as rural since they are in the three least urban categories: small metropolitan, micropolitan, and noncore.

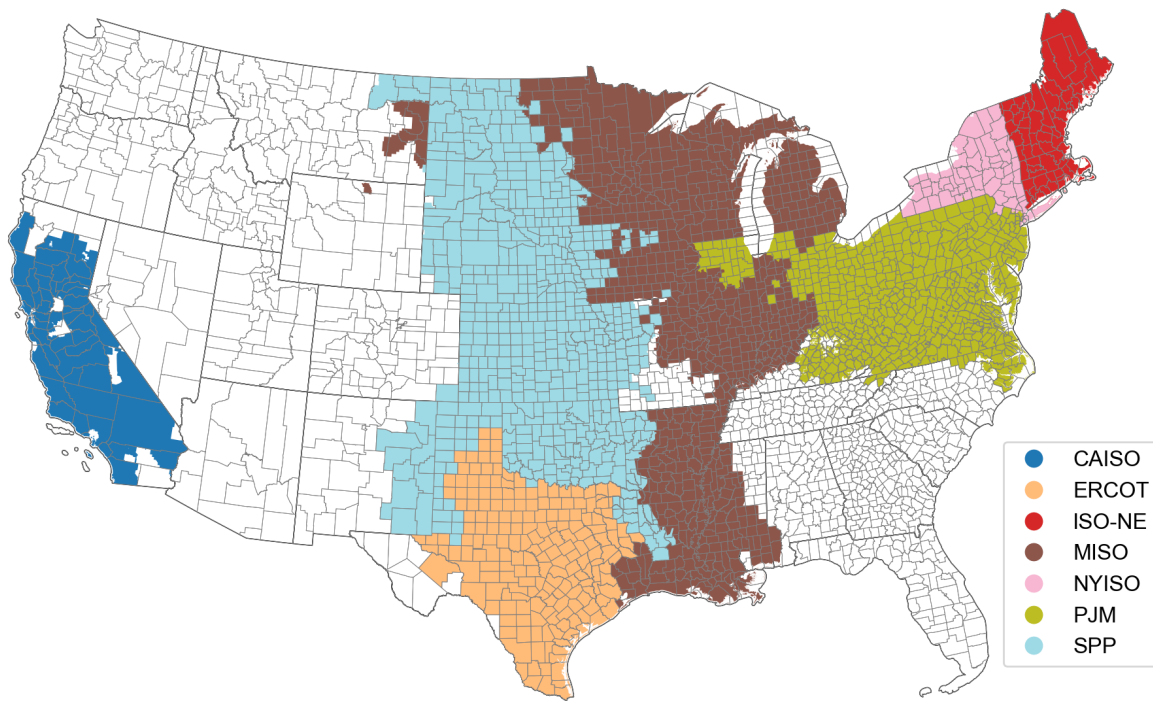

**Fig 7. Spatial distribution of US Regional Transmission Organizations (RTOs).** The US has seven RTOs, where utilities and other high-voltage owners pool their transmission assets to enable greater efficiency over a large network[2]. These RTOs includes California ISO (CAISO), Southwest Power Pool (SPP), Electric Reliability Council of Texas (ERCOT), Midcontinent ISO (MISO), New York ISO (NYISO), New England ISO (ISO-NE) and PJM.

**Table 1. Multicollinearity test result of the features using the variance inflation factor (VIF).**

| Features | VIF |
|---|---|
| Number of events | 8.5638 |
| Average outage rate | 1.9239 |
| Average duration | 4.0043 |
| Average inter-event time | 1.6012 |
| Cumulative number of customers affected | 1.9610 |
| Peak number of customers affected | 7.9097 |
| Average increase/decrease rate | 1.7139 |
| Average duration per customer experienced | 1.8773 |
| Number of events affecting >5% customers | 3.8854 |
| Average duration affecting >5% customers | 1.5119 |
| Average inter-event time affecting >5% customers | 1.7242 |
| Number of events exceeding 12 hours | 4.1140 |
| Average outage rate exceeding 12 hours | 4.6381 |
| Average inter-event time exceeding 12 hours | 1.5912 |

**Table 2. The tuned best hyperparameters for the XGBoost model.**

| Hyperparameters | Values |
|---|---|
| gamma | 0.1507 |
| learning_rate | 0.0646 |
| max_depth | 9 |
| min_child_weight | 2 |
| n_estimators | 139 |
| reg_lambda | 1.0 |
| subsample | 0.7128 |

**Table 3. Performance of machine learning models.** We evaluated nine widely used classification models. The F1 score is served as the primary performance indicator for model selection. Among the models, XGBoost achieved the highest F1 score.

| Models | F1-score | AUC | Accuracy | Precision | Recall |
|---|---|---|---|---|---|
| XGBoost | 0.7937 | 0.8955 | 0.7835 | 0.8051 | 0.7826 |
| Logistic Regression | 0.7477 | 0.8203 | 0.7340 | 0.7921 | 0.7080 |
| Decision Tree | 0.7548 | 0.7549 | 0.7488 | 0.7852 | 0.7267 |
| Random Forest | 0.7892 | 0.8914 | 0.7802 | 0.8058 | 0.7733 |
| SVM | 0.7437 | 0.8382 | 0.7488 | 0.8605 | 0.6549 |
| Gradient Boosting | 0.7844 | 0.8859 | 0.7802 | 0.8203 | 0.7516 |
| AdaBoost | 0.7781 | 0.8704 | 0.7653 | 0.7830 | 0.7733 |
| LightGBM | 0.7876 | 0.8886 | 0.7736 | 0.7864 | 0.7888 |
| CatBoost | 0.7886 | 0.8897 | 0.7802 | 0.8078 | 0.7702 |

**Note. Description of the Urban/rural form and structure features**

These features are from our previous work[3]. We collected a diverse range of features through systematic literature review to capture various heterogenous features related to urban/rural form and structure. Urban/rural form and structure are concepts in urban planning and geography that describe the physical layout and organization of areas[4-6]. Due to data unavailability, our analysis of the Pearson correlation between urban/rural form and structure dimensions and the PSVI includes a total of 867 rural counties and 1658 urban counties.

**GDP:** To estimate the status of the economic development of the county, we adopted the 2019 data of gross domestic product for each county. The data are provided by the Bureau of Economic Analysis in the US Department of Commerce[7].

**Population density:** The population size was obtained from the 2020 race and ethnicity data from US Census Bureau[8]. We calculated the population density at the county level by dividing the total population of the county by its land area. Land area data was also obtained from US Census Bureau[9].

**Minority segregation and income segregation:** Segregation refers to the physical and social separation of different racial, ethnic, and socioeconomic groups[10]. This separation can take many forms, including minority segregation and income segregation. One of the key consequences of segregation is that it often leads to unequal distribution of resources, as well as increased exposure to environmental hazards[11,12]. We adopted the Dissimilarity Index (DI) to evaluate minority segregation and income segregation. The DI is a measure of spatial segregation that indicates the extent to which two groups are evenly distributed across different areas, which ranges from 0 (indicating perfect evenness) to 1 (indicating complete separation)[13,14]. We calculated the DI based on the proportion of minority population (for minority segregation) and the proportion of low-income population (for income segregation) at the census-tract level relative to the county level[15]:

$$DI = \frac{1}{2} \sum_{i=1}^{n} \left| \frac{x_i}{X} - \frac{y_i}{Y} \right| \qquad (1)$$

where, $x_i$ is the minority population (or low-income population) in the census tract unit; $X$ is the the minority population (or low-income population) in the county unit; $y_i$ is the reference population in the census tract unit; $Y$ is the reference population in the county unit.

For minority segregation, we collected the racial population data from the 2020 race and ethnicity data from US Census Bureau[8]. The primary racial groups are non-Hispanic White, non-Hispanic Black, and non- Hispanic Asian residents. Non-Hispanic populations are selected because White, Black, or Asian populations can be mutually selective from Hispanic populations. We considered the non-Hispanic Black, and non- Hispanic Asian as the minority population and non-Hispanic White as the reference population.

For income segregation, we extracted median income data from the 2020 American Community Survey (ACS)[16]. We used the 5-year estimates of median income due to the broader coverage of areas, larger sample size, and higher precision, making the data more reliable than 1-year and 3-year estimates. We used the quantile income groups of a county (Q1 to Q4) to indicate income levels with Q1/Q2 representing low-income groups and Q3/Q4 represent high-income groups, respectively.

**POI density:** To capture the distribution of physical facilities, we adopted the 6.5 million active POI data in the US from SafeGraph[17]. The dataset includes basic information about POIs, such as POI IDs, location names, geographical coordinates, addresses, brands, and North American Industry Classification System (NAICS) codes to categorize POIs. The NAICS code is the standard used by federal statistical agencies in classifying business establishments[18]. We selected 10 essential types of POIs that are closely relevant to human daily lives: restaurants, schools, grocery stores, churches, gas stations, pharmacies and drug stores, banks, hospitals, parks, and shopping malls. We counted the number of POIs in each county and calculated its density as their facility distribution feature.

**Road density:** To capture the distribution of transportation network, we extracted data from Open Street Map[19] to calculate the density of road segments in counties. We estimated complete road networks from the raw data by assembling road segments. Since the lengths of road segments created by the source were in close proximity, we calculated road density by dividing the number of road segments by the areas of a county.

**Urban centrality index:** We adopted urban centrality index (UCI) to characterize the centralization degree of the facilities in a county. UCI is the product of the local coefficient and the proximity index[20]. The local coefficient was computed based on the number of POIs within each census tract; the proximity index was computed based on the number of POIs within each census tract along with a distance matrix that considered the distance between census tracts. The value of UCI ranges from 0 to 1. The values close to 0 indicate polycentric distribution of facilities within a county, while the values close to 1 indicate monocentric distribution of facilities. The indices are formulated as follows[20]:

$$LC = \frac{1}{2}\sum_{i=1}^{N}(k_i - \frac{1}{N}) \quad (2)$$

$$PI = 1 - \frac{V}{V_{max}} \quad (3)$$

$$V = K' \times D \times K \quad (4)$$

where, $N$ is the total number of census tracts in a county; $K$ is a vector of the number of POIs in each census tract; $k_i$ is a component of the vector $K$; $D$ is the distance matrix between census tracts; $V_{max}$ is calculated by assuming that the total POIs are uniformly settling on the boundary of the county; $LC$ is the local coefficient, which measures the unequal distribution; $PI$ is the proximity index, which resolves the normalization issue; $V$ is the Venables Index.

**Human mobility index:** To understand the inequality of population activities, we employed mobile phone data from Spectus Inc. to develop the metric of human mobility index (HMI). The data has a wide set of attributes, including anonymized user ID, latitude, longitude, POI ID, time of observation, and the dwelling time of each visit[21]. We extracted the data from April 2019 (28 days) to account for the variation of population activities on weekdays and weekends. Our period is also during regular conditions when no external extreme events perturbed human activities. To develop the HMI, we first assigned each visit point $v_i$ to a defined CBG in a county. Then, we calculated HMI as follows:

$$HMI = \frac{\sum_{i=1}^{n} v_i}{28n} \tag{5}$$

where *n* denotes the number of CBGs in a county.

We mapped the values of HMI to the range from 0 to 1 using min-max scaling. The proximity of HMI values to 0 or 1 indicates the level of human mobility and activity, with values closer to 0 indicating lower activity and values closer to 1 indicating higher activity in a county.

To reduce complexity while preserving the essential information of the features, we implemented principal component analysis (PCA), a statistical technique used for dimensionality reduction[22], to the eight features to identify the most important components of urban/rural form and structure. The best number of principal components was selected as three, and the cumulative explained variance of 90.59% indicates that these three principal components capture a considerable amount of the variability in the original data and provide a meaningful representation of the urban/rural form and structure.

The first principal component is named development density (DD), which includes the features of population density, POI density, and road density, explaining 33.41% of the total variance. This component represents the level of urbanization and built environment density in a given area. The second component is defined as centrality and segregation (CS), explaining 27.56% of the total variance and including the features of UCI, minority segregation, and income segregation. This component represents the level of social and economic segregation, as well as the degree of urban centralization in a given area. The third component, economic activity (EA), explains 29.62% of the total variance and includes the features of GDP and HMI. This component represents the level of economic activity and mobility in a given area.